\documentclass[aps,superscriptaddress,showpacs,pre,10pt,footinbib,twocolumn]{revtex4-2}
\usepackage{bm} 
\usepackage{graphicx} 
\usepackage{amsmath}
\usepackage{amsthm}
\usepackage{amssymb} 
\usepackage{epstopdf} 
\usepackage{comment}
\usepackage{amsfonts}
\usepackage{epsfig}
\usepackage{color}
\usepackage{algorithm}
\usepackage{algpseudocode}
\usepackage[dvipsnames]{xcolor}
\usepackage{subfigure}
\usepackage[english]{babel}
\usepackage[bb=boondox]{mathalfa}
\usepackage{mathtools}
\usepackage{dsfont}
\usepackage{tabularx}
\usepackage{booktabs}

\usepackage{hyperref}
\usepackage[normalem]{ulem} 
\hypersetup{
    colorlinks=true,       
    linkcolor=cyan,          
    citecolor=magenta,        
    filecolor=magenta,      
    urlcolor=cyan,           
    runcolor=cyan
}
\newcommand{\ket}[1]{| #1 \rangle}

\newcommand {\be}{\begin{equation}}
\newcommand {\ee}{\end{equation}}

\newcommand{\ba}{\begin{eqnarray}}
\newcommand{\ea}{\end{eqnarray}}
\newcommand\tr{{\mbox{Tr\,}}}
\newcommand{\ignore}[1]{}
\newcommand{\avg}[1]{\left< #1 \right>}

\newcommand\ident{{\mathbb{1}}}

\usepackage[margin=1in,nofoot]{geometry}

\newcommand{\Siq}{S_{{\bf{i}}_q}}
\newcommand{\e}{{{e}}}

\renewcommand{\Re}{\operatorname{Re}}

\newcommand{\beq}{\begin{equation}}
\newcommand{\eeq}{\end{equation}}
\newcommand{\beqnn}{\begin{equation*}}
\newcommand{\eeqnn}{\end{equation*}}
\newcommand{\bea}{\begin{eqnarray}}
\newcommand{\eea}{\end{eqnarray}}
\newcommand{\beann}{\begin{eqnarray*}}
\newcommand{\eeann}{\end{eqnarray*}}
\newcommand{\bes} {\begin{subequations}}
\newcommand{\ees} {\end{subequations}}

\newcommand{\sgn}{\textrm{sgn}}

\begin{document}
\raggedbottom 
\title{A quantum Monte Carlo algorithm for arbitrary high-spin Hamiltonians}
\author{Arman Babakhani}
\affiliation{Information Sciences Institute, University of Southern California, Marina del Rey, California 90292, USA}
\affiliation{Department of Physics and Astronomy, University of Southern California, Los Angeles, California 90089, USA}
\author{Lev Barash}
\affiliation{Information Sciences Institute, University of Southern California, Marina del Rey, California 90292, USA}
\author{Itay Hen}
\email{itayhen@isi.edu}
\affiliation{Information Sciences Institute, University of Southern California, Marina del Rey, California 90292, USA}
\affiliation{Department of Physics and Astronomy, University of Southern California, Los Angeles, California 90089, USA}

\date{\today}

\begin{abstract}
\noindent
We present a universal quantum Monte Carlo algorithm for simulating 
arbitrary high-spin (spin greater than $1/2$) Hamiltonians, 
based on the recently developed permutation matrix representation (PMR) framework.
Our approach extends a previously developed PMR-QMC method for spin-$1/2$ Hamiltonians 
[Phys. Rev. Research 6, 013281 (2024)].
Because it does not rely on a local bond decomposition, the method applies equally well 
to models with arbitrary connectivities, long-range and multi-spin interactions, 
and its closed-walk formulation allows a natural analysis of sign-problem conditions in terms of cycle weights.
To demonstrate its applicability and versatility, we apply our method to spin-$1$ and spin-$3/2$
quantum Heisenberg models on the square lattice, as well as to randomly generated high-spin Hamiltonians.
Additionally, we show how the approach naturally extends to general Hamiltonians involving 
mixtures of particle species, including bosons and fermions. 
We have made our program code freely accessible on GitHub.
\end{abstract}

\maketitle

\section{Introduction}

Quantum high-spin models, namely quantum many-body Hamiltonians depicting the interactions between particles possessing spins greater than $1/2$, are pivotal for understanding complex magnetic interactions, quantum phase transitions, topological phases, and quantum entanglement in condensed matter systems~\cite{Haldane1983,Reja2024,Sandvik_2010}. Such models provide a rich framework for both theoretical studies and experimental realizations, enabling the exploration of new physical phenomena that are not present in simpler spin-$1/2$ models including quantum phase transitions, topological order, entanglement, and exotic magnetic states. Some notable examples are the high-spin Heisenberg models, Haldane chains which include an exchange interaction and a single-ion anisotropy term, 
and the Affleck-Kennedy-Lieb-Tasaki (AKLT) model, which serves as an example of a spin-1 chain with a ground state that exhibits a Haldane gap and non-trivial topological order~\cite{Affleck1988}.

In the literature, a variety of approaches have been proposed to tackle numerous large-scale high-spin models, ranging from a functional renormalization group approach to study Heisenberg models with unrestricted spin length~\cite{Baez_2017}
to a high-order coupled cluster method applied to Kagome lattice antiferromagnets with arbitrary spin restricted to translationally invariant lattices~\cite{G_tze_2011}.
Also notable are studies investigating spin-liquid behavior in a spin-one Kitaev model under magnetic fields~\cite{Zhu_2020} which are closely tied to the unique solvability of the Kitaev model. 

The main workhorse in the study of high-spin quantum systems is quantum Monte Carlo (QMC) techniques. 
The stochastic series expansion (SSE) with operator-loop updates~\cite{Sandvik_1999} has been successful for simulating Heisenberg models. 
Directed loop algorithms~\cite{Syljuasen2002} extended the SSE approach providing improved ergodicity and efficiency, albeit their implementation for high spins remain Hamiltonians remain model-specific. Additionally, there have been numerous QMC proposals to address the sign-problem for frustrated spin systems. For example, in Ref.~\cite{Wojtkiewicz_2007} a QMC scheme for frustrated Heisenberg antiferromagnets has been developed to mitigate the sign-problem in frustrated systems.
Other studies (see, e.g., Ref.~\cite{Alvarez_2002}) explored mixed-spin quantum magnets where spin magnitudes vary across sites, and the finite-temperature behavior of square-lattice spin-one Heisenberg antiferromagnets~\cite{Harada_1998}.

As the above literature survey suggests, high-spin models are of significant relevance in condensed matter physics; however, existing numerical techniques for simulating high-spin quantum many-body systems remain constrained in both generality and applicability. In particular, QMC algorithms often rely on update schemes that are intricately tailored to the specific symmetries and properties of the model under study, rendering them non-transferable across different systems and limiting their broader utility.

This work aims to remedy this situation by providing in contrast a universal framework for studying arbitrarily complex high-spin quantum models of any spin value and interactions of essentially any geometry, dimension, locality, and connectivity. Furthermore, a natural generalization of the proposed framework to address QMC simulations of mixed-species models including spin-$1/2$ particles, bosons, and fermions is also discussed in detail. 

The present technique builds on a recent study by the authors~\cite{PMRQMC2024} in which a universal QMC algorithm designed to simulate arbitrary spin-$1/2$ Hamiltonians was devised. There, it was shown that 
casting the to-be-simulated Hamiltonian in Permutation Matrix Representation (PMR) form~\cite{pmr} allows one to write the partition function of any spin-$1/2$ system as a sum of efficiently calculable terms each of which is associated with a closed walk on the computational state graph of the Hamiltonian \footnote{The computational state graph of a Hamiltonian is the graph $G_H = (V,E)$ whose nodes are the computational basis states, and the edges (connectivity and weights) are given by the off-diagonal terms of the Hamiltonian $H$.} and where Monte Carlo updates to faithfully sample these walks may be automatically generated in a systematic way~\cite{pmr,PMRQMC2024}.  

The PMR-QMC framework has already been benchmarked in several
spin-$1/2$ settings, including transverse-field Ising and XXZ models, the toric code, and random MAX2SAT-type
Hamiltonians~\cite{ODE,pmr,PMRQMC2024}. In these studies PMR-QMC was found to equilibrate reliably, and in
some nonlocal problems, such as random MAX2SAT in a transverse field, it outperformed optimized SSE
implementations by orders of magnitude in wall-clock time while yielding results that agree within
statistical error.

In this paper, we generalize the aforementioned technique to the case of high-spin (spin greater than $1/2$) Hamiltonians and introduce a similar-in-spirit universal Monte Carlo algorithm designed to reliably simulate arbitrary high-spin systems. To that aim, we devise a protocol for generating the necessary set of QMC updates, based on the PMR decomposition, that ensure an ergodic Markov chain Monte Carlo sampling of the partition function of essentially any conceivable input system. We illustrate that while for spin-$1/2$ systems achieving the same goal required finding the nullspace of sets of binary (modulo 2) vectors representing the permutation matrices of the Hamiltonian, for spin-$s$ particles the task is generalized to finding a similar set of modulo $(2s+1)$ vectors. 

 The paper is organized as follows. In Sec.~\ref{sec:Hamiltonians}, we provide a brief overview of the permutation matrix representation quantum Monte Carlo (PMR-QMC) and analyze high-spin Hamiltonians in this context. In Sec.~\ref{sec:qmc}, we discuss the QMC algorithm, describing in detail the method we have devised to generate all the necessary QMC updates and demonstrating how these moves ensure both ergodicity and detailed balance. There, we also discuss the emergence of the sign problem in our scheme. 
In Sec.~\ref{sec:Results} we showcase the power of our technique by presenting simulation results for two models, namely, the spin-$1$ and spin-$3/2$ quantum Heisenberg models on the square lattice, as well as for randomly generated spin-$1$ and spin-$3/2$ Hamiltonians. In Sec.~\ref{sec:qmcext} we examine in detail how the approach taken can be extended to include other particle species as well as mixtures thereof. We conclude in Sec.~\ref{sec:conclusions} with an additional discussion and future directions of research. 

\section{Permutation matrix representation for high-spin Hamiltonians}\label{sec:Hamiltonians}

\subsection{Overview of PMR-QMC}\label{sec:perMatRep}

We begin by providing a brief overview of the permutation matrix representation (PMR) protocol~\cite{pmr,PMRQMC2024} on which the simulation algorithm will be based. PMR begins by first casting the to-be-simulated Hamiltonian 
in PMR form, i.e., as a sum
\beq \label{eq:basic}
H=D_0 + \sum_{j=1}^M D_j P_j \,,
\eeq
where $D_j$ are diagonal matrices and 
$\{P_j\}_{j=1}^M$ are permutation matrices. 
As shown in~\cite{pmrAdvanced}, these permutation matrices can be chosen as a subset of a special Abelian group $G$~\footnote{Every $P \in G$ 
is a permutation matrix with no nonzero diagonal elements, except for the identity matrix $P_0 = \ident$. 
For every $\ket{i}, \ket{j}$ pair, there exists a unique $P \in G$ such that $P\ket{i} = \ket{j}$. 
See \cite{pmrAdvanced} for more details.}. 
For spin-1/2 systems, $G$ consists of all Pauli-X strings~\cite{PMRQMC2024}. For $S>\frac{1}{2}$ systems, we will describe a decomposition of spin operators ($SU(2)$ generators), in section \ref{subsec:PMR_highspin}, using permutation matrices that are by construction Abelian.

A non-trivial, though useful, consequence is that we can write the partition function 
as a sum of `generalized Boltzmann 
weights'~\cite{pmr, ODE, PMRQMC2024} 
\beq
\mathcal{Z} = \sum_{z}\sum_{q=0}^\infty\sum_{S_{{\bf{i}}_q}=\mathbb{1}}  D_{(z,S_{{\bf{i}}_q})}   \e^{-\beta [E_{z_0},\ldots,E_{z_q}]} = \sum_{\cal C} w_{\mathcal C},
\label{eq:PartitionFunction}
\eeq
where each $\mathcal{C} = (z, \Siq)$ is a QMC configuration, 
and each weight 
\beq\label{eq:wc}
w_{\cal C} = D_{(z, \Siq)} e^{-\beta [E_{z_0}, \ldots, E_{z_q}]}
\eeq
is efficiently computable.
Here, \hbox{${\bf{i}}_q=(i_1,\ldots,i_q)$} is a multi-index, where each index $i_j$ runs from $1$ to $M$, and $\Siq \equiv P_{i_q}\ldots P_{i_1}$ denotes a product of $q$ permutations, 
each from $\{P_j\}_{j=1}^M$. 
The summation over $\cal C$ is a double 
sum over all basis states $\ket{z}$ and all possible products $\Siq$ that evaluate to the identity, 
for $q$ from $0$ to $\infty$. 
Next, we denote $\ket{z_0}\equiv \ket{z}$ and $\ket{z_k} \equiv P_{\mathbf{i}_k} \ldots P_{\mathbf{i}_1} \ket{z}$
for $k = 1,2,\dots, q$.
This notation allows us to define the `diagonal-energies' as $E_{z_k} \equiv \avg{z_k | H | z_k} = \avg{z_k| D_0 | z_k}$ 
and the off-diagonal `hopping strength,' $D_{(z, \Siq)} \equiv \prod_{k=1}^q d^{({\bf{i}}_k)}_{z_k}$,
where $d^{({\bf{i}}_k)}_{z_k} = \avg{z_k | D_{i_k} | z_k}$.
Finally, $e^{-\beta [E_{z_0}, \ldots, E_{z_q}]}$ denotes the divided difference~\cite{deboor:05,pmr} of $f(x) = e^{-\beta x}$ with respect to the inputs 
$\{E_{z_0}, \ldots, E_{z_q}\}$, 
which can be efficiently computed in practice in $O(q)$ time~\cite{GuptaBarashHen2020}.

At the formal level, as expressed in Eqs.~(\ref{eq:PartitionFunction}) and (\ref{eq:wc}),
the PMR-QMC framework can be viewed as a reorganization of the conventional stochastic series expansion (SSE) representation.
Like Handscomb's original diagrammatic expansions~\cite{handscomb1,handscomb2} and Sandvik's SSE algorithm~\cite{Sandvik_1999}, 
PMR-QMC starts from the Taylor expansion of $e^{-\beta H}$.
However, instead of sampling explicit operator strings built from local bond terms, PMR-QMC groups the Taylor terms into closed walks of basis states 
and expresses their weights as divided differences of the exponential.
In this representation PMR-QMC is no longer a high-temperature expansion, and a single PMR-QMC configuration aggregates an entire family of SSE operator strings that differ only in the ordering and multiplicity of diagonal insertions.

The validity of the PMR construction does not depend on a particular choice of computational basis: 
given any orthonormal basis of the Hilbert space, one can construct the corresponding permutation matrices 
and run PMR-QMC in that basis. Although the explicit forms of the matrices change from one basis to another, 
the structure of the algorithm and its Monte Carlo updates remains the same. This basis flexibility is a 
useful practical feature, as it allows one to perform local basis rotations which, in some cases, 
can mitigate or even remove a sign problem by mapping the Hamiltonian into a more favourable form 
(cf.~Sec.~\ref{sec:signProb}).

To construct a Markov chain, it is necessary to employ real-valued non-negative weights,
whether or not a sign problem~\cite{vgp} is present.
To this end, one can consider either absolute values of real components of the weights or absolute values of the weights:
\beq
W_{{\cal C}}^{(1)} = \left| \Re[w_{\cal C}] \right|,\quad W_{{\cal C}}^{(2)} = \left| w_{\cal C} \right|\,.
\label{eq:gbw}
\eeq

The PMR formulation allows one to measure a wide range of static operators and dynamical quantities~\cite{pmrAdvanced}. 
The key to being able to do so is to write for any given operator $A$ its thermal average as
\beq
\langle A\rangle = \frac{\tr[A\e^{-\beta H}]}{\tr[\e^{-\beta H}]}=\frac{\sum_{\cal C} A_{\cal C}w_{\cal C}}{\sum_{\cal C}w_{\cal C} }.
\label{eq:meanA}
\eeq
Although, generally, both $w_{\cal C}$ and $A_{\cal C}$ are complex-valued,
both sums $\sum_{\cal C} A_{\cal C}w_{\cal C}$ and $\sum_{\cal C}w_{\cal C}$ are real-valued
since both $H$ and $A$ are Hermitian operators. Therefore, we have
\bea
\label{eq:A}
\langle A\rangle &=&
\frac{\sum_{\cal C} \left(\Re[A_{\cal C}w_{\cal C}]/W_{\cal C}^{(1)}\right)\cdot W_{\cal C}^{(1)}}{\langle\sgn\rangle_1\sum_{\cal C} W_{\cal C}^{(1)}},\\
\langle A\rangle &=&
\frac{\sum_{\cal C} \left|A_{\cal C}\right| \cos(\arg(A_{\cal C}w_{\cal C})) \cdot W_{\cal C}^{(2)}}{\langle\sgn\rangle_2\sum_{\cal C} W_{\cal C}^{(2)}},
\eea
where
$\sgn_1(w_{\cal C}) = \sgn(\Re[w_{\cal C}])$,
$\sgn_2(w_{\cal C}) = \cos(\arg(w_{\cal C}))$, and
\bea
\langle\sgn\rangle_1 &=&
\frac{\sum_{\cal C} \sgn_1(w_C) \cdot W_{\cal C}^{(1)}}{\sum_C W_{\cal C}^{(1)}},\\
\langle\sgn\rangle_2 &=&
\frac{\sum_{\cal C} \sgn_2(w_C) \cdot W_{\cal C}^{(2)}}{\sum_C W_{\cal C}^{(2)}}.
\eea
The values of $\Re[A_{\cal C} w_{\cal C}]/W_{\cal C}^{(1)}$ and
$\left|A_{\cal C}\right| \cos(\arg(A_{\cal C}w_{\cal C}))$,
therefore, represent the instantaneous quantity associated with the configuration ${\cal C} = {(z,{S_{{\bf{i}}_q}})}$
that will be collected during the simulation
when using the weights $W_{\cal C}^{(1)}$ and $W_{\cal C}^{(2)}$, respectively. 
For further details about calculating a wide range of observables throughout the simulation,
see Refs.~\cite{pmr,PMRQMC2024,pmrAdvanced,ezzell2025universalblackboxquantummonte}.
In particular, this formalism allows one to construct estimators for essentially arbitrary static operators
as well as for imaginary-time correlation functions and their integrated susceptibilities.
The explicit examples of such constructions given in Refs.~\cite{pmrAdvanced,ezzell2025universalblackboxquantummonte}
are fully applicable to the high-spin case.

\subsection{PMR decomposition of single spin operators} \label{subsec:PMR_highspin}

Before discussing the PMR formulation of general high-spin Hamiltonians, let us first briefly review high-spin operators.
The matrix elements of the
$(2s+1) \times (2s+1)$ spin operators $X$, $Y$, and $Z$ for
\hbox{$s \in \{1/2, 1, 3/2, 2, 5/2, \ldots\}$} are given by
\bea
  &X_{jk} & = \frac{1}{2}  \left(\delta_{j,k+1} + \delta_{j+1,k}\right) \sqrt{(s + 1)(j + k - 1) - jk}, \nonumber\\
  &Y_{jk} & = \frac{j}{2} \left(\delta_{j,k+1} - \delta_{j+1,k}\right) \sqrt{(s + 1)(j + k - 1) - jk}, \nonumber\\
  &Z_{jk} & = (s + 1 - j) \delta_{j,k}, \nonumber
\eea
where $j, k \in \{1,2,\dots,2s+1\}$.

In order to devise a PMR decomposition for these spin operators, let us define the following $(2s+1) \times (2s+1)$ permutation matrix
\beq
                         P = 
                           \begin{pmatrix}
                             0 & 1 & 0 & \ldots & 0\\
                             0 & 0 & 1 & \ldots & 0\\
                             \vdots & \vdots & \ddots & \ddots & 0\\
                             0 & 0 & \ldots & 0&  1 \\
                             1 & 0 & \ldots & 0&  0 \\
                           \end{pmatrix}.
\label{eq:Pmatrix}
\eeq
We must note that $P^{2s+1}= \mathbb{1}$ and that $P$ has no fixed points on the computational basis states (eigenstates of the spin-$Z$ operator).
Using this permutation matrix, we can write the spin operators as follows
\bea
&X &= D^{+} P + D^{-} P^{-1}, \label{eq:X} \\
&Y &= - i D^{+} P + i D^{-} P^{-1}, \label{eq:Y} \\
&Z &= D^{(z)} \,, \label{eq:Z}
\eea
where $D^+$, $D^-$, $D^{(z)}$ are diagonal matrices with the following diagonal entries:
\bea
&D^{(z)}_{j} &= s+1-j, \,  \\
&D^{+}_{j}&=\frac{1}{2} \,\sqrt{(2 s-j+1)j}, \,  \\
&D^{-}_{j}&=\frac{1}{2} \,\sqrt{(2(s+1)-j)(j-1)}, 
\eea
with $j=1, \ldots, 2s+1$.

For reasons that will become clear in the following sections, we also introduce the matrices $D^{(k)}$ and $D^{(z,k)}$ such that
\bea
D^{(k)}_{j} &=& D^+_{(j+k)\bmod{(2s+1)}},\\
D^{(z,k)}_{j} &=& D^{(z)}_{(j+k)\bmod{(2s+1)}},
\eea
where $D^+_0$ and $D^{(z)}_0$ are defined as $D^+_0 = D^+_{2s+1} = 0$ and $D^{(z)}_0 = D^{(z)}_{2s+1} = -s$.
We note that $D^+ = D^{(0)}$, $D^- = D^{(-1)}$, and $D^{(z)} = D^{(z,0)}$.
The matrices $D^{(k)}$ and $D^{(z,k)}$ obey
\bea
P D^{(k)} &=& D^{(k+1)} P, \label{eq:PD+} \\
P^{-1} D^{(k)} &=& D^{(k-1)} P^{-1}, \label{eq:PD-} \\
P D^{(z,k)} &=& D^{(z,k+1)} P, \label{eq:PDz+} \\
P^{-1} D^{(z,k)} &=& D^{(z,k-1)} P^{-1}. \label{eq:PDz-}
\eea

We highlight the features of this PMR decomposition through the following example. Consider the following two-spin operator $Y \otimes X$ and let the spin $s>1/2$. The PMR decomposition for this will be
\begin{align}
    Y \otimes X &= (- i D^{+} P + i D^{-} P^{-1}) \otimes (D^{+} P + D^{-} P^{-1}) \notag \\
    &= -i ( D^+ P \otimes D^+ P + D^+ P \otimes D^- P^{-1}) \notag  \\
    &\,\,\,\,\, +i(D^- P^{-1} \otimes D^+ P + D^- P^{-1} \otimes D^- P^{-1}) \notag \\
    & = -i\Big( (D^+\otimes D^+) P_1 + (D^+\otimes D^-) P_2 \notag \\
    &\quad  -(D^-\otimes D^+) P_3 - (D^-\otimes D^-) P_4\Big) \,,
\end{align}
where $P_1 = P \otimes P$, $P_2 = P \otimes P^{-1}$, $P_3 = P^{-1} \otimes P$, and $P_4 = P^{-1} \otimes P^{-1}$. We emphasize that for spin-$1/2$, $P=P^{-1}=X$, so there will only be a single unique permutation operator as $P_1 = P_2 = P_3 = P_4$. However, this is not that case for higher spin ($s>1/2$) systems. Additionally, one can verify that these permutation operators commute, i.e. $[P_i , P_j] = 0$ for $i,j\in \{1,2,3,4\}$, highlighting the fact that tensor product of powers of single-spin permutation operators $P$ are Abelian and can be used for PMR decomposition for any spin Hamiltonian.
\subsection{High-spin Hamiltonians}

Consider now an $n$-particle Hamiltonian given as the linear combination 
\beq\label{eq:HH1}
H = \sum_i c_i \prod_{k=1}^{m_i} s^{(i)}_{j_{i,k}},
\eeq
where $c_i$ are real-valued coefficients and $\prod_{k=1}^{m_i} s^{(i)}_{j_{i,k}}$ are spin operator strings. 
Here, $s^{(i)}_{j_{i,k}}$ represents a spin matrix $s \in \{X,Y,Z\}$
in the $i$-th string acting on the $j_{i,k}$-th particle, where $j_{i,k}\in\{1,2,\dots,n\}$.
The operator $s^{(i)}_{j_{i,k}}$ is a tensor product of a spin matrix and $n-1$ identity matrices such that
it has the same matrix dimension as the Hamiltonian.
Each spin operator string may contain any number of any of the three spin matrices for each of the particles in any order.

Equation~(\ref{eq:HH1}) can be rewritten as
\beq\label{eq:HH}
H = \sum_i c_i \bigotimes_{j=1}^n \prod_{k=1}^{m_{i,j}} s^{(i)}_{j,k},
\eeq
where $s^{(i)}_{j,k} \in \{X,Y,Z\}$ denotes a spin matrix 
in the $i$-th term acting on the $j$-th particle, and $m_{i,j}$ are non-negative integers.

We will now cast the general Hamiltonian $H$, Eq.~(\ref{eq:HH}),
in PMR form. To do that, we apply Eqs.~(\ref{eq:X})--(\ref{eq:Z}) to each $s^{(i)}_{j,k}$ and rewrite the Hamiltonian in the form
\beq
H = \sum_i d_i \bigotimes_{j=1}^n \prod_{k=1}^{m_{i,j}} t^{(i)}_{j,k},
\eeq
where each $t^{(i)}_{j,k}$ acts on the $j$-th particle
and is either $D^{(0)} P$ or $D^{(-1)} P^{-1}$ or $D^{(z,0)}$. 

Next, we `push' all diagonal matrices  in each product $\prod_{k=1}^{m_{i,j}} t^{(i)}_{j,k}$ to the left
using the relations~(\ref{eq:PD+})--(\ref{eq:PDz-}) to obtain
\beq
H = \sum_i d_i \bigotimes_{j=1}^n D^{(i,j)} \bigotimes_{j=1}^n P^{(i,j)},
\label{eq:PMR1}
\eeq
where $D^{(i,j)}$ and $P^{(i,j)}$ are diagonal and permutation matrices, respectively, and
the index $j$ indicates the action on the $j$-th particle.
The matrix $P^{(i,j)}$ is equal to the matrix $P$ raised to some power $n_{i,j} \in \{0,1,\dots,2s\}$.

Last, we group together all terms that have the same $\bigotimes_{j=1}^n P^{(i,j)}$ component,
ending up with a Hamiltonian of the form
\beq
H = \sum_i D_i P_i,
\label{eq:PMR2}
\eeq
where $P_i = \bigotimes_{j=1}^n P^{n_{i,j}}$.

We have thus achieved a PMR decomposition for arbitrary spin-$s$ Hamiltonians. 

\section{The QMC algorithm}\label{sec:qmc}

\subsection{QMC configurations} 

For any Hamiltonian cast in PMR form, the partition function $Z=\tr[e^{-\beta H}]$ can be written as a sum of configuration weights [cf.~Eq.~(\ref{eq:PartitionFunction})], where a configuration \hbox{${\cal C}=\{|z\rangle, S_{{\bf{i}}_q}\}$} is a pair of 
a classical (diagonal) basis state $|z\rangle$ and a product $S_{{\bf{i}}_q}$ of permutation operators that must evaluate to the identity element $P_0=\mathbb{1}$.
The configuration ${\cal C}$ induces a list of states \hbox{$\{|z_0\rangle =|z\rangle ,|z_1\rangle ,\ldots,|z_q\rangle =|z\rangle \}$}, which in turn generates a corresponding multi-set of
energies $E_{{\cal C}}=\{E_{z_0},E_{z_1},\ldots,E_{z_q}\}$ for the configuration.

We can now consider a QMC algorithm that samples these configurations with probabilities proportional to their weights $W_{{\cal C}}$, Eq.~(\ref{eq:gbw}). The Markov process would start with some initial configuration and a set of (probabilistic) rules, or QMC updates, will dictate transitions from one configuration to the next. 

We will take the initial state to be 
\hbox{$\mathcal{C}_0=\{|z\rangle, S_0=\mathbb{1}\}$} where $|z\rangle$ is a randomly generated initial classical state.
The weight of this initial configuration is 
\beq
W_{{\cal C}_0}=\e^{-\beta [E_z]}=\e^{-\beta E_z} \,,
\eeq
i.e., the classical Boltzmann weight of the initial randomly generated basis state $|z\rangle$. 

The set of required QMC updates will be discussed in the next sections. To ensure that configurations are sampled properly, i.e. in proportion to their weight, one must ensure that the Markov process is ergodic, i.e., that the QMC updates are capable of generating all basis states $|z\rangle$ as well as all sequences $S_{{\bf{i}}_q}$ evaluating to the identity. An additional sufficient requirement to ensure proper sampling is that of detailed balance, which dictates that the ratio of transition probabilities from one configuration to another and the transition in the opposite direction equals to the ratio of their respective weights~\cite{bookNewmanBarkema,bookGrimmett}.
In the following, we show how both conditions are made to be satisfied for general high-spin Hamiltonians with arbitrary interactions.

\subsection{Fundamental cycles}
\label{sec:fundamentalcycles}

It follows from Eqs.~(\ref{eq:PMR1}) and (\ref{eq:PMR2}) that the permutation operators $P_i$ are tensor products of powers of $P$.
As such, (i) all permutation operators commute, and (ii) $P^{2s+1}=\mathbb{1}$, so each permutation operator
$P_i = \bigotimes_{j=1}^n P^{n_{i,j}}$ 
satisfies $P_i^{2s+1}=\mathbb{1}$. 

Denoting by $p_i$ the integer-string $[n_{i,1} n_{i,2} \cdots n_{i,n}]$, one can easily verify
that the product of two permutation operators $P_i$ and $P_k$ would likewise correspond to modular 
addition of the components of $p_i$ and $p_k$ modulo $2s+1$, i.e. 
\beq
P_i P_k \rightarrow p_i + p_j \mod 2s+1 \, .
\eeq
For a sequence of operators evaluating to the identity,
$S_{{\bf{i}}_q}=P_{i_q} \ldots P_{i_1}$, 
we have $\sum_{j=1}^q p_{i_j} \equiv {\bf 0}\, (\text{mod } 2s+1)$,
where ${\bf 0}$ is an integer-string consisting of only zeros.
We note that $S_{{\bf{i}}_q}$ is a permutation of the multiset of operators $\{P_{i_1}, \dots, P_{i_q}\}$,
which can be represented as an integer string $[a_1 \, a_2 \, \dots \, a_M]$,
where $a_k$ denotes the number of occurrences of the operator $P_k$ among $P_{i_1}, \dots, P_{i_q}$.

The question of how one can generate all possible sequences of operators (which evaluate to the identity) 
is therefore reduced to the question of how one can generate all integer-strings 
$[a_1 \, a_2 \, \dots \, a_M]$ that obey the following system of linear equations over mod-$(2s+1)$ addition
\beq
\sum_{i=1}^M a_i \cdot n_{i,j} \equiv 0 \pmod{2s+1}, \, j = 1,2,\dots,n.
\label{eq:mod2s1}
\eeq
Equation~(\ref{eq:mod2s1}) can be solved by finding the null space basis over addition modulo $2s+1$ for the matrix  
${\bf n}^T$ whose columns are $p_1^T, \ldots, p_M^T$.
For details on accomplishing this task regardless of the decomposition of $2s+1$ into prime factors, we utilize the fact that a system of modular equivalences Eq.~(\ref{eq:mod2s1}) can be rewritten as a system of Diophantine equations
\beq
\sum_{i=1}^M a_i \cdot n_{i,j} + (2s_j+1)\cdot k_j = 0, \, j = 1,2,\dots,n,
\label{app:dioph}
\eeq
where $k_1, k_2,\dots, k_n$ are additional unknown integers, and $s_1=\dots=s_n=s$.
The Diophantine system Eq.~(\ref{app:dioph}) can be solved by employing Hermite normal form 
of the corresponding matrix. For details, the reader is referred to Refs.~\cite{schrijver1998book,Cohen1993book}.

We shall call a multiset of permutation operators that multiply to the identity a `cycle'.
The length of a cycle would be the number of distinct permutation operators in it.
We shall refer to cycles represented by the integer strings from the null space basis as fundamental cycles.

Generally, the nullspace basis states can be chosen in many different ways, and so the choice of the set of fundamental cycles is not unique.
From a practical point of view, however, we find that obtaining a `minimal cycle basis', i.e., a basis that minimizes the lengths 
of all basis cycles, is advantageous from the QMC standpoint.
This follows from the fact that the probability of a QMC update to be accepted is a decreasing function of the cycle length. 
To reduce the cycles lengths, we therefore find a null space basis and then proceed to replace long-cycle basis states 
with shorter basis states by performing mod-$(2s+1)$ additions between the integer-strings of pairs of cycles,
accepting the changes each time a new cycle with a shorter length is found. 
The process ends when a pass through all pairs of cycles does not result in an improvement.

It follows from the properties of null space that any cycle can be obtained by (i) insertion and removal of fundamental cycles,
(ii) insertion and removal of trivial cycles consisting of $2s+1$ identical permutation operators, and
(iii) swapping the order of two adjacent permutation operators.

\begin{table*}[thb]
\setlength{\tabcolsep}{6pt}
\centering
\begin{tabularx}{\textwidth}{lXX}
\toprule
Update & Spin-$1/2$ QMC simulations & High-spin QMC simulations \\
\midrule
Pair insertion and deletion 
&
The pair consists of two identical permutation operators.
&
The pair consists of two mutually inverse permutation operators, which can be either the same or different.
\\\midrule
Classical update 
& 
The update performs a spin-flip of a random spin.
& 
The update selects a random particle and changes its spin value to a randomly selected different value.
\\\midrule
Fundamental cycle completion 
&
The update that consists of choosing a subsequence $S$ from $S_{{\bf{i}}_q}$,
choosing a fundamental cycle containing all operators of the subsequence $S$, and
attempting to replace the subsequence $S$ with the remaining operators from the selected cycle.
A subsequence $S$ must not contain repeated operators. The elements of $S$ are not required
to be consecutive within $S_{{\mathbf i}_q}$.
& 
If $A$ and $B$ are two complementary parts of a cycle, we do not directly replace $A$ with $B$.
Instead, we replace either $A^{-1}$ by $B$ or $A$ by $B^{-1}$.
The selected subsequence $S$ in $S_{{\mathbf i}_q}$ may now contain repetitions,
and the expressions for the acceptance probability are adjusted accordingly.
See details in Appendix~\ref{appendix:cycle_completion}.
\\\midrule
Worm update 
&
The update is rejected with a small probability $p_f$ at each intermediate step,
where $p_f$ is an adjustable parameter.
Intermediate configurations are assigned their `natural' weight $W_{\cal C}$ 
as per Eq.~(\ref{eq:gbw}).
& 
We find it useful not to artificially exit the worm, i.e., we use $p_f = 0$.
To prevent the worm from straying too far from being healed, each intermediate configuration $\cal C$ 
is assigned the weight $W_{\cal C} \exp(-\alpha d)$.
where $W_{\cal C}$ is defined by Eq.~(\ref{eq:gbw}), $\alpha$ 
is an adjustable parameter that may depend on system size and temperature, and $d$ is the `distance from identity,' 
defined as the number of particles with different spin values in \( |z_0\rangle \) and \( |z_q\rangle \).

\\
\bottomrule
\end{tabularx}
\caption{Main differences in QMC updates between the spin-$1/2$ case (covered in Ref.~\cite{PMRQMC2024})
and the high-spin case considered in the present work.}
\label{tab:QMC}
\end{table*}

\subsection{QMC updates}
\label{subsec:QMCupdates}

As a preliminary step, prior to the simulation taking place, we carry out the PMR decomposition of the Hamiltonian,
and produce a list of fundamental cycles for the to-be-simulated Hamiltonian cast in PMR form (see details
in Sections above).
Additionally, we add $M$ trivial cycles to the list of fundamental cycles, each containing $2s+1$ identical permutation operators.

Because the Hamiltonian is Hermitian, for each operator $P_i$ from the PMR decomposition~(\ref{eq:basic}), 
the operator $P_i^{-1}$ is also included in the PMR.
Hence, the list of permutation operators consists of pairs of mutually inverse operators, 
as well as the operators that are inverse to themselves.

We employ the following QMC updates: (i) the worm update, (ii) block swap, and (iii) classical updates.

The worm update involves single operator moves performing a `disturbance' 
causing $S_{{{\bf i}}_q}$ to evaluate to a non-identity permutation,
as well as `healing' back to an identity-forming sequence (see details in Ref.~\cite{PMRQMC2024}).
Within the worm update, we also employ local swaps, pair insertions and deletions, 
and fundamental cycle completions. 

The basic QMC updates mentioned above are similar in nature to analogous moves used in the spin-$1/2$ case~\cite{PMRQMC2024},
with a few necessary important changes which are summarized in Table~\ref{tab:QMC}.
In particular, fulfillment of the detailed balance condition can be shown similarly to the spin-$1/2$ case~\cite{PMRQMC2024}.

It was shown in the previous section that local swaps, fundamental cycle completions, and trivial cycle completions 
are sufficient to ensure ergodicity along the quantum (or imaginary-time) dimension, 
i.e., the ability to generate all permutation operator sequences that evaluate to the identity.
Also, we find applying the worm update very useful in practice because it further accelerates 
Markov chain mixing and the achievement of ergodicity in the quantum dimension.

The ergodicity along the classical dimension, i.e., the generation of all possible classical basis states $|z\rangle$,
is achieved by employing the classical update.

For two arbitrary configurations ${\cal C} = \{ |z\rangle, S_{{\bf{i}}_q} \}$ and ${\cal C'} = \{ |z'\rangle, S_{{\bf{i'}}_{q'}} \}$
such that $W_{\cal C} \ne 0$ and $W_{\cal C'} \ne 0$,
the above QMC updates allow in particular the following sequence of transformations:
${\cal C} \to {\cal C}_0 \to {\cal C}'_0 \to {\cal C}'$, where
${\cal C}_0 = \{ |z\rangle, \mathbb{1} \}$ and ${\cal C}'_0 = \{ |z'\rangle,\mathbb{1}\}$.
Therefore, the transformation from ${\cal C}$ to ${\cal C}'$ is possible, and the ergodicity 
holds in the entire configuration space.

In our implementation, almost all of the computational cost is due to computing and updating the divided differences that define the configuration weights.
If $q$ denotes the number of energies along a sampled walk (which scales proportionally to $\beta N$) and $s$ the total dimensionless energy spread along the walk
($s \propto \max_{i,j} \lvert \beta E_i - \beta E_j \rvert$), then updating the divided difference after a generic local change of the walk requires $O(q s)$ arithmetic operations~\cite{GuptaBarashHen2020}.
In the parameter regimes relevant for our benchmarks, both $q$ and $s$ grow approximately linearly with $\beta N$, leading to a per-update cost of order $O(\beta^2 N^2)$.

The classical update, which regenerates the classical configuration $\ket{z}$, is the only move that requires a full recomputation of the divided differences, at cost $O(q^2 s)$.
We therefore employ the classical update only with a modest frequency in order to control this overhead.
Since another move (the block swap) also changes the classical state, the classical update can be used rather infrequently 
while still ensuring ergodicity along the classical dimension.

The cost of the worm update is proportional to the typical worm length, which is controlled by the tuning parameter $\alpha$ (see Table~\ref{tab:QMC}).
Very small values of $\alpha$ lead to excessively long worms and a large runtime, whereas excessively large values of $\alpha$ cause the worm to terminate too quickly and reduce its effectiveness as an ergodicity-enhancing update that reduces autocorrelation times.
In practice, we choose $\alpha$ such that typical worm lengths remain moderate and hence do not exhibit a clear polynomial growth with $\beta$ or $N$, so that the leading $O(\beta^2 N^2)$ scaling is not altered and the dependence on $\alpha$ enters only through the prefactor.

To further reduce autocorrelation times, especially at larger $\beta$, we have also implemented two additional Monte Carlo moves:
inner-loop reversal moves, which reverse the orientation of a closed loop of permutations while keeping its cycle structure unchanged,
and loop-exchange moves, which interchange two inner loops, i.e., swap their positions in the operator sequence 
without modifying their internal structure. 
Both moves respect detailed balance, and we have found them to improve the convergence of the simulations 
without changing the basic $O(\beta^2 N^2)$ per-update scaling.

\subsection{Emergence of the sign problem in high-spin PMR-QMC}\label{sec:signProb}

As was detailed in previous sections, the PMR-QMC method prescribes a specific decomposition of the partition function of input high-spin Hamiltonians, casting it as a sum of efficiently computable weights. Based on those, a set of QMC update rules are generated that are shown to guarantee the convergence of the Monte Carlo Markov chain to its proper thermal distribution, by ensuring the ergodicity of the Markov chain all the while satisfying detailed balance via importance sampling. 

Since Markov chain Monte Carlo requires treating the weights as (unnormalized) probabilities, whenever the decomposition produces negative weights, the algorithm encounters what is commonly referred to as the sign problem, as was discussed in Sec.~\ref{sec:perMatRep}. As was also noted above, in the presence of a sign problem, one must resort to sampling the QMC configurations with respect to modified weights that are guaranteed to be positive. The price one has to pay for sampling from an incorrect distribution is often exponentially longer convergence times. 

For this reason, the conditions under which PMR-QMC encounters a sign problem are of interest. We thus turn next to analyzing the emergence of the sign problem in PMR-QMC.   
Examining the condition for the positivity of the weight $w_{\cal{C}}$ given in Eq.~(\ref{eq:wc}), we first note that the term $e^{-\beta[E_{z_0},\ldots,E_{z_q}]}$ is positive (negative) for even (odd) values of $q$ \footnote{This is a general property of the divided difference for the function $f(x) = e^{-\beta x}$. Check Ref. \cite{vgp} for details.}, the length of the walk and so the sign of a summand can be simplified to
\beq \label{eq:PosCond}
\sgn \left[W_{(z,\Siq)}\right]=\sgn \Re\left[\prod_{k=1}^q (-d^{({\bf{i}}_k)}_{z_k})\right],
\eeq
i.e., the PMR expansion of the partition function will admit a negative weight if and only if there exists a closed walk on the computational state graph (the weighted graph whose adjacency matrix is the Hamiltonian) along which $\Re\left[\prod_{k=1}^q (-d^{({\bf{i}}_k)}_{z_k})\right]<0$ (the reader is referred to Ref.~\cite{vgp} for additional details). A necessary and sufficient condition for a sign-problem-free Hamiltonian is thus that all the complex phases of the cycles of the computational state graph are zero (modulo $2 \pi$); a condition that we refer to as VGP (for `vanishing geometric phase'). Explicitly, the condition means that all `cycles' of the computational state graph, namely, the products of matrix elements of the form $(-H_{ij})(-H_{jk})(-H_{km})\cdots(-H_{\ell i})$ must have zero complex phase ($\text{mod }2\pi$). 

It is worth noting that it is sometimes the case that different flavors of QMC, which prescribe different decompositions of the partition function, may differ in the severity and in some cases also in the mere presence of the sign problem.   One notable example is QMC schemes in which the Hamiltonian is written as a sum of local bonds (such as SSE~\cite{sandvik:92,sandvik:03}) wherein the condition for a sign-problem-free simulation is more stringent than that of PMR-QMC, namely, while PMR-QMC requires that products of off-diagonal matrix elements must be non-negative, the SSE condition is that all products of \emph{local} bond strengths must be non-negative.


Despite the enhanced capabilities of PMR-QMC, systems that are truly afflicted by a severe sign problem in the VGP sense remain challenging in practice.
However, the VGP condition~\cite{vgp} significantly enlarges the sign-problem-free domain beyond the usual stoquastic class.
In particular, any Hamiltonian of the form
\[
H_{\mathrm{VGP}} = U H_{\mathrm{stoq}} U^\dagger,
\]
where $H_{\mathrm{stoq}}$ is stoquastic in some local basis and $U$ is a diagonal unitary, is generically nonstoquastic in the computational basis but remains sign-problem free in the PMR-QMC formulation.
A simple family of examples is obtained by starting from a stoquastic transverse-field Ising or XXZ Hamiltonian and applying arbitrary local $Z$-rotations to each spin.
In the rotated computational basis the resulting model is in general no longer stoquastic, so a straightforward SSE representation in that basis would suffer from a sign problem, whereas the model remains in the VGP class and can be simulated sign-problem free within the PMR-QMC framework.
Further nonstoquastic yet sign-problem-free spin models of this type, including flux-attached versions of the toric code and certain frustrated Ising models with complex couplings related by diagonal gauge transformations to stoquastic ones, have been constructed explicitly in Ref.~\cite{vgp}.
These classes provide examples of physical systems where the PMR-QMC framework can treat models that would be inaccessible, or at least severely hampered by a sign problem, in standard SSE or world-line formulations.

\section{Results}\label{sec:Results}

In this section, we illustrate the effectiveness and scope of PMR-QMC in studying a variety high-spin Hamiltonians. In the next subsection, we provide simulation results for spin-$1$ and spin-$3/2$ Heisenberg models. The subsection that follows is dedicated to the study of random Hamiltonians, which existing methods are ill posed to simulate. 

Wherever exact calculations were possible, we verified the correctness and accuracy of our technique 
by ensuring that the calculated values agree with the exact results.
In particular, we performed verification for a dozen small systems with fewer than ten spins 
in each of the cases $s = 1$, $s = 3/2$, $s = 2$, and $s = 5/2$.

\subsection{High-spin quantum Heisenberg models}
\label{subsec:Heisenberg}

We begin by probing the thermodynamic behavior of the high-spin quantum Heisenberg model 
on a square $L\times L$ lattice with open boundary conditions under an external magnetic field. 
There is no algorithmic restriction to open boundaries, and periodic boundary conditions can be implemented just as easily by
including the corresponding couplings in the Hamiltonian input; the PMR-QMC algorithm itself is unchanged by this choice.
The Hamiltonian of the model is given by
\beq
H=-J\sum_{\langle i,j\rangle}\left(X_i X_j + Y_i Y_j+  Z_i Z_j \right)+h\sum_{i=1}^n Z_i\,,  
\eeq
where $n = L^2$ is the number of particles, $\langle i,j\rangle$ denotes neighbors on the lattice, and we choose for our simulations the interaction strength to be $J=1$ and the external magnetic field strength to be $h=0.1$.

For these chosen parameters, we compute via our QMC algorithm the specific heat
\beq
C = \beta^2 \left( \langle E^2\rangle - \langle E\rangle^2 \right),
\eeq
and the magnetic susceptibility
\beq
\chi_{\cal M} = \beta \left( \langle {\cal M}^2\rangle - \langle {\cal M}\rangle^2  \right),
\eeq
where above $\cal M$ is the $Z$-magnetization \hbox{${\cal M} = \frac{1}{s N} \sum_{i=1}^N Z_i$}. 
We have computed the dependence 
of both quantities on the inverse temperature $\beta$
across various system sizes. 

\begin{figure}[t]
\includegraphics[width=\columnwidth]{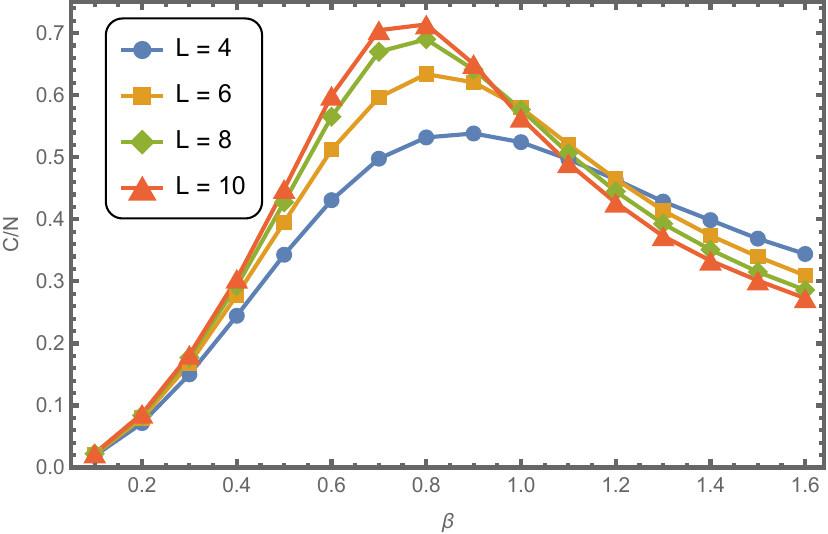}
\includegraphics[width=\columnwidth]{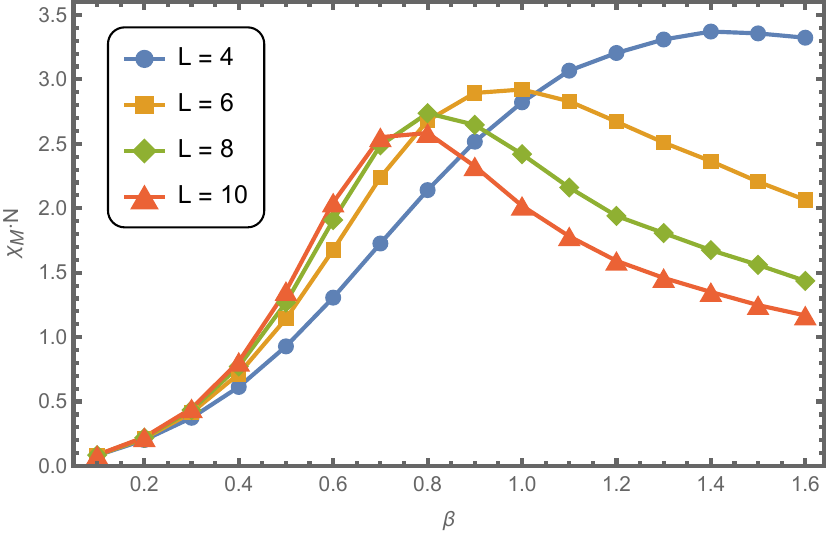}
\caption{
Calculations of the  spin-$1$ quantum Heisenberg model on a square $L\times L$ lattice. Top: Specific heat as a function of inverse-temperature $\beta$.
Bottom: magnetic susceptibility as a function of $\beta$. 
The standard errors are no larger than the size of a marker.
}
\label{fig:QH3}
\end{figure}
\begin{figure}[t]
\includegraphics[width=\columnwidth]{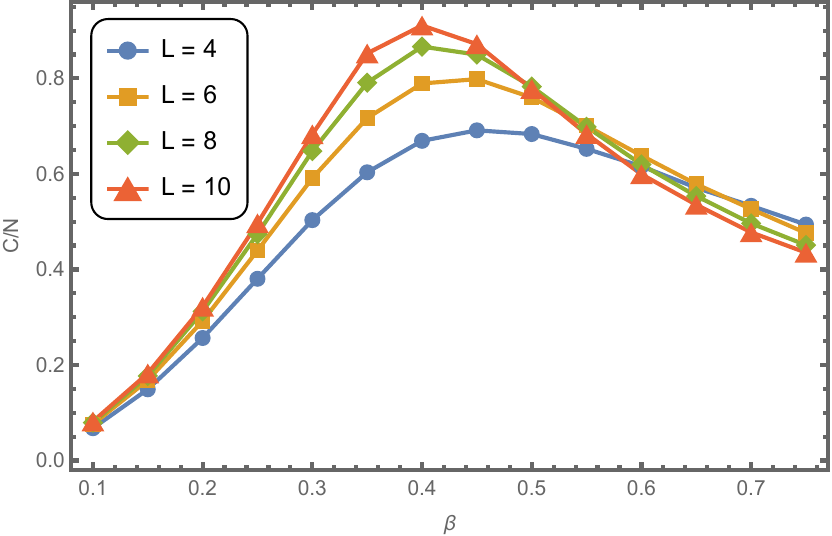}
\includegraphics[width=\columnwidth]{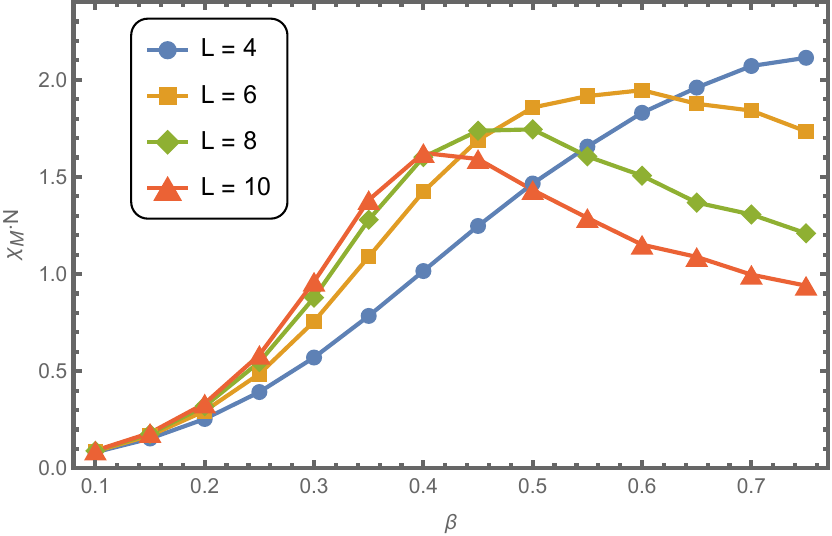}
\caption{
Calculations of the  spin-$3/2$ quantum Heisenberg model on a square $L\times L$ lattice. Top: Specific heat as a function of inverse-temperature $\beta$.
Bottom: magnetic susceptibility as a function of $\beta$.
The standard errors are no larger than the size of a marker.
}
\label{fig:QH4}
\end{figure}

In Fig.~\ref{fig:QH3}, we present results for the spin-$1$ quantum Heisenberg model, 
while Fig.~\ref{fig:QH4} shows corresponding data for systems with spin magnitude $s=3/2$. 
Both figures display the specific heat and magnetic susceptibility 
as functions of inverse temperature, calculated for varying lattice sizes 
under a small external magnetic field.

A prominent feature observed in both cases is the presence of peaks in specific heat 
and magnetic susceptibility. However, these peaks do not signify conventional phase transitions 
characterized by spontaneous symmetry breaking and sharp critical behavior. 
Instead, they correspond to a thermal crossover phenomenon, marking the gradual evolution 
from quantum-dominated behavior at lower temperatures to classical paramagnetic behavior at high temperatures, 
under the influence of a small external magnetic field.

This behavior is expected for the two-dimensional quantum Heisenberg model in a small external magnetic field. 
In the absence of long-range order and without a diverging correlation length 
at finite temperature (as constrained by the Mermin--Wagner theorem in the zero-field limit), 
the system cannot undergo a true phase transition at finite temperature. 
The external field explicitly breaks the SU(2) symmetry, further suppressing the possibility 
of spontaneous symmetry breaking. As a result, a thermal crossover behavior is expected
rather than a finite-temperature phase transition.

\begin{table*}[t]
\small
\begin{tabular}{|c|c||c|c|c|c|c|c|c|c|c|c|c|c|c|c|c|c|}
\hline
 spin-$1$ & $\beta:$ &  $0.1$ & $0.2$ & $0.3$ & $0.4$ & $0.5$ & $0.6$ & $0.7$ &  $0.8$ & $0.9$ & $1.0$ & $1.1$ & $1.2$ & $1.3$ & $1.4$ & $1.5$ & $1.6$ \\
\hline
\hline
$L=4$ & $\alpha:$  &   $1$     &  $1$     &   $1$    &   $1$    & $1$   &    $1$   &  $1$    &  $1$     &   $1$   &  $1$    &  $1$  &  $1$  &  $1.5$  & $1.5$  &  $1.5$   & $1.5$ \\
      & $N_{updates}:$  &   $10^8$     &  $10^8$     &   $10^8$    &   $10^8$    & $10^8$   &    $10^8$   &  $10^8$    &  $10^8$     &   $10^8$   &  $10^8$    &  $2\cdot 10^8$  &  $2\cdot 10^8$  &  $4\cdot 10^8$  & $8\cdot 10^8$  &  $8\cdot 10^8$   & $8\cdot 10^8$ \\
      & $N_{cpus}:$ &  20          &    20       &    20       &    20       &   20     &   20        &  20        &   20        &    20      &    20      &   20     &    20    &  20      &   20    &  20       &  20    \\
      & \footnotesize \verb~CPU time~: & \footnotesize   402        & \footnotesize   1197      & \footnotesize   1284      & \footnotesize    3153     & \footnotesize   6566   & \footnotesize  12616      & \footnotesize  41837     & \footnotesize  71144   & \footnotesize   105418   & \footnotesize   107567   & \footnotesize   290008    & \footnotesize 388420     & \footnotesize 144824     & \footnotesize 323860    & \footnotesize  357050   & \footnotesize  386193 \\
\hline
$L=6$ & $\alpha:$  &   $2.5$     &  $2.5$     &   $2.5$    &   $2.5$    & $2.5$   &    $2.5$   &  $2.5$    &  $2.5$     &   $2.5$    &  $2.5$   &  $2.5$   &   $2.5$    &   $2.5$    &  $2.5$    &  $2.5$   &   $2.5$    \\
      & $N_{updates}:$  &   $10^8$     &  $10^8$     &   $10^8$    &   $10^8$    & $10^8$   &    $10^8$   &  $10^8$    &  $10^8$     &   $10^8$    &  $10^8$   &  $10^8$   &   $10^8$    &   $10^8$    &  $10^8$    &  $10^8$   &   $10^8$    \\
      & $N_{cpus}:$ &       20     &   20        &   20        &     20      &   20     &    20       &    20      &    20       &    20      &     20     &  100     &  100     &  100     &  100    &   100     &  100   \\
      & \footnotesize \verb~CPU time~: & \footnotesize    338       & \footnotesize  843        & \footnotesize   1968      & \footnotesize  2595       & \footnotesize  6200    & \footnotesize  8432       & \footnotesize 14888    & \footnotesize   19844     & \footnotesize 20707      & \footnotesize    27080   & \footnotesize   24813  & \footnotesize   27778  & \footnotesize  47951   & \footnotesize  51882  & \footnotesize   63688   & \footnotesize  65330 \\
\hline
$L=8$ & $\alpha:$  &   $3$     &  $3$     &   $3$    &   $3$    & $3$   &    $3$   &  $3$    &  $3$     &   $3$    &  $3$   &  $3$   &   $3$    &   $3$    &  $3$    &  $3$   &   $3$    \\
      & $N_{updates}:$  &   $10^8$     &  $10^8$     &   $10^8$    &   $10^8$    & $10^8$   &    $10^8$   &  $10^8$    &  $10^8$     &   $10^8$    &  $10^8$   &  $10^8$   &   $10^8$    &   $10^8$    &  $2\cdot 10^8$    &  $2\cdot 10^8$   &   $2\cdot 10^8$    \\
      & $N_{cpus}:$ &       20     &   20        &   20        &     20      &   20     &    20       &    20      &    50       &    60      &    160     &  200     &  200     &  200     &  200    &   180     &  200   \\
      & \footnotesize \verb~CPU time~: & \footnotesize 644          & \footnotesize 1142        & \footnotesize 1744        & \footnotesize 6771        & \footnotesize 12372    & \footnotesize 19798       & \footnotesize 27153      & \footnotesize 53966       & \footnotesize 68130      & \footnotesize 89907      & \footnotesize 102265   & \footnotesize  78168   & \footnotesize  139456  & \footnotesize  322720 & \footnotesize  270480   & \footnotesize 323285 \\
\hline
$L=10$ & $\alpha:$ &   $3$     &  $3$     &   $3$    &   $3$    & $3$   &    $3$   &  $3$    &  $3$     &   $3.5$    &  $3.5$   &  $3.5$   &   $3.5$    &   $3.5$    &  $3.5$    &  $3.5$   &   $3.8$    \\
       & $N_{updates}:$ &   $10^8$     &  $10^8$     &   $10^8$    &   $10^8$    & $10^8$   &    $10^8$   &  $10^8$    &  $10^8$     &   $2\cdot 10^8$    &  $4\cdot 10^8$   &  $4\cdot 10^8$   &   $4\cdot 10^8$    &   $4\cdot 10^8$    &  $8\cdot 10^8$    &  $8\cdot 10^8$   &   $1.6\cdot 10^9$    \\
      & $N_{cpus}:$ &       20     &   20        &   20        &     20      &   50     &    50       &    50      &    50       &    200     &     200    &  200     &  200     &  200     &  200    &   200     &  200   \\
      & \footnotesize \verb~CPU time~: & \footnotesize 2321         & \footnotesize  4514       & \footnotesize 8183        & \footnotesize 16115       & \footnotesize 27415    & \footnotesize 65285       & \footnotesize 118253     & \footnotesize 189086      & \footnotesize 156645     & \footnotesize 287518     & \footnotesize 542418   & \footnotesize 461619   & \footnotesize 667700   & \footnotesize  711055 & \footnotesize 738242    & \footnotesize 1925783       \\
\hline
\end{tabular}
\caption{Calculation parameters and computational effort required for spin-$1$ quantum Heisenberg model simulations.}
\label{tab:params_2S+1=3}
\end{table*}
\begin{table*}[t]
\small
\begin{tabular}{|c|c||c|c|c|c|c|c|c|c|c|c|c|c|c|c|}
\hline
 spin-$3/2$ & $\beta:$ & $0.10$ & $0.15$ & $0.20$ & $0.25$ & $0.30$ & $0.35$ & $0.40$ &  $0.45$ & $0.50$ & $0.55$ & $0.60$ & $0.65$ & $0.70$ & $0.75$ \\ 
\hline
\hline
$L=4$ & $\alpha:$  &   $1$     &  $1$     &   $1$    &   $1$    & $1$   &    $1$   &  $1$    &  $1$     &   $1$   &  $1.5$    &  $1.5$  &  $1.5$  &  $1.5$  & $1.5$  \\ 
      & $N_{updates}:$  &   $10^8$     &  $10^8$     &   $10^8$    &   $10^8$    & $10^8$   &    $10^8$   &  $10^8$    &  $10^8$     &   $10^8$   &  $10^8$    &  $10^8$  &  $10^8$  &  $10^8$  & $10^8$  \\ 
      & $N_{cpus}:$ &  20          &    20       &    20       &    20       &   20     &   20        &  20        &   20        &    20      &    20      &   20     &    20    &  20      &   20    \\ 
      & \footnotesize \verb~CPU time~: & \footnotesize 686          & \footnotesize 3195        & \footnotesize 5003        & \footnotesize 11733       & \footnotesize 35131    & \footnotesize 58300       & \footnotesize 158345     & \footnotesize 345201      & \footnotesize 848711     & \footnotesize   81475    & \footnotesize  104902  & \footnotesize 135302   & \footnotesize 159605   & \footnotesize  187511 \\ 
\hline
$L=6$ & $\alpha:$  &   $2.5$     &  $2.5$     &   $2.5$    &   $2.5$    & $2.5$   &    $2.5$   &  $2.5$    &  $2.5$     &   $2.5$    &  $2.5$   &  $2.5$   &   $2.5$    &   $2.5$    &  $2.5$    \\ 
      & $N_{updates}:$  &   $10^8$     &  $10^8$     &   $10^8$    &   $10^8$    & $10^8$   &    $10^8$   &  $10^8$    &  $10^8$     &   $10^8$    &  $10^8$   &  $10^8$   &   $10^8$    &   $10^8$    &  $10^8$    \\ 
      & $N_{cpus}:$ &  20          &    20       &    70       &    20       &   200    &   200       &  200       &   200       &    200     &    200     &   200    &    200   &  200     &   200   \\ 
      & \footnotesize \verb~CPU time~: & \footnotesize 744          & \footnotesize 1127        & \footnotesize 2979        & \footnotesize  3099       & \footnotesize 9527     & \footnotesize 19232       & \footnotesize 29037      & \footnotesize 60088       & \footnotesize 58977      & \footnotesize 154829     & \footnotesize 173850   & \footnotesize 190066   & \footnotesize 241744   & \footnotesize  171422 \\ 
\hline
$L=8$ & $\alpha:$  &   $3$     &  $3$     &   $3$    &   $3$    & $3$   &    $3$   &  $3$    &  $3$     &   $3$    &  $3$   &  $3$   &   $3$    &   $3.5$    &  $3.5$    \\ 
      & $N_{updates}:$  &   $10^8$     &  $10^8$     &   $10^8$    &   $10^8$    & $10^8$   &    $10^8$   &  $10^8$    &  $10^8$     &   $10^8$    &  $10^8$   &  $10^8$   &   $10^8$    &   $10^8$    &  $10^8$    \\ 
      & $N_{cpus}:$ &  20          &    20       &    20       &    20       &   20     &   80        &  80        &   140        &    170      &   170      &   170    &   198    &  200     &   200   \\ 
      & \footnotesize \verb~CPU time~: & \footnotesize 1132         & \footnotesize  1745       & \footnotesize  7714       & \footnotesize  7312       & \footnotesize 14625    & \footnotesize  33025      & \footnotesize 61143      & \footnotesize 141865      & \footnotesize 214170     & \footnotesize 235157     & \footnotesize 364372   & \footnotesize 320661   & \footnotesize 112047   & \footnotesize 85999   \\ 
\hline
$L=10$ & $\alpha:$ &   $3.5$     &  $3.5$     &   $3.5$    &   $3.5$    & $3.5$   &    $3.5$   &  $3.5$    &  $3.5$     &   $3.5$    &  $3.5$   &  $3.5$   &   $3.5$    &   $3.8$    &  $3.8$    \\ 
       & $N_{updates}:$ &   $10^8$     &  $10^8$     &   $10^8$    &   $10^8$    & $10^8$   &    $10^8$   &  $10^8$    &  $10^8$     &   $10^8$    &  $10^8$   &  $10^8$   &   $10^8$    &   $2\cdot 10^8$    &  $4\cdot 10^8$    \\ 
      & $N_{cpus}:$ &  40          &    40       &    40       &    40       &   40     &   40        &  160       &   200       &    200     &    200     &   200    &    200   &  200     &   200   \\ 
      & \footnotesize \verb~CPU time~: & \footnotesize 1782         & \footnotesize 3826        & \footnotesize 7312        & \footnotesize 11521       & \footnotesize 36812    & \footnotesize 75154       & \footnotesize 99435      & \footnotesize 123472      & \footnotesize 123222     & \footnotesize 251064     & \footnotesize 245735   & \footnotesize 274114   & \footnotesize 536974   & \footnotesize 622954  \\ 
\hline
\end{tabular}
\caption{Calculation parameters and computational effort required for spin-$3/2$ quantum Heisenberg model simulations.}
\label{tab:params_2S+1=4}
\end{table*}

This crossover originates from the competition between quantum fluctuations, dominant at lower temperatures, 
and thermal fluctuations, which become increasingly significant as temperature rises. 
At low temperatures (large $\beta$), quantum fluctuations stabilize correlated quantum states. 
The presence of the external magnetic field additionally breaks rotational symmetry explicitly, 
influencing the magnetization response and subtly modifying the energy landscape, 
thus affecting the crossover behavior. 
At higher temperatures (small $\beta$), thermal excitations dominate, disrupting quantum correlations 
and steering the system toward classical paramagnetism. 
Consequently, the observed peaks reflect enhanced fluctuations around the temperature range 
where neither quantum nor classical fluctuations overwhelmingly dominate.

Comparing spin-$1$ and spin-$3/2$ models, clear differences in thermodynamic behavior are apparent. 
For the spin-$3/2$ system, peaks appear at smaller inverse temperatures (higher absolute temperatures) 
compared to the spin-$1$ case. This shift is explained by the larger separation between energy levels 
in higher-spin systems, which effectively reduces their susceptibility to thermal excitations. 
Consequently, the transition toward classical paramagnetic behavior occurs at higher temperatures.

In addition, higher-spin particles are expected to exhibit reduced quantum fluctuations,
leading to a more classical behavior at low temperatures.
The peaks in the spin-$3/2$ data are sharper and better-defined, indicative of reduced quantum fluctuations. 
In contrast, the spin-$1$ system exhibits broader and less distinct peaks, signifying 
stronger quantum fluctuations that extend the temperature interval of the crossover, 
blurring the transition between quantum and classical regimes.

These numerical observations underscore the nuanced thermodynamic behavior of high-spin quantum Heisenberg models, 
highlighting how spin magnitude significantly influences the interplay between quantum and thermal fluctuations. 

To complete the picture, Tables~\ref{tab:params_2S+1=3} and~\ref{tab:params_2S+1=4} show the number of QMC updates for each simulation, $N_{updates}$,
and the parameter $\alpha$ used in the calculations (see Sec.~\ref{subsec:QMCupdates}).
The parameters for each calculation were chosen to ensure that the autocorrelation diagnostics were satisfied 
and that the error bars did not exceed the marker size in Figures~\ref{fig:QH3} and~\ref{fig:QH4}.

The PMR-QMC algorithm is particularly amenable to parallel execution, with each independently run Markov chain 
providing an equal contribution to the accumulated statistics.
In particular, we observe a near-perfect strong scaling speedup 
similar to that observed in the spin-$1/2$ case (see Sec.~VI-F in Ref.~\cite{PMRQMC2024}).

The use of parallelization allows one to carry out autocorrelation diagnostics for the algorithm, recalling that statistical errors are commonly obtained from binning analysis for standard observables, and from jackknife analysis for derived observables such as specific heat and magnetization (see, e.g., Refs.~\cite{Janke2008,PMRQMC2024}).
In a parallelized calculation, one may therefore compare the statistical error obtained from a single run 
with the error estimated from independent -- and therefore uncorrelated -- parallel executions of the algorithm.
A satisfactory agreement between these two estimates would suggest that the measurement blocks are effectively uncorrelated, implying that the autocorrelation time is shorter than the block length. 
If the two estimates differ, albeit not substantially, this typically suggests that the autocorrelation time 
exceeds the block length, yet remains much shorter than the timescale of the entire simulation.
If, however, the error obtained from binning analysis is significantly underestimated compared to the statistical error 
estimated from independent runs (e.g., by a factor of two or more), this typically indicates that 
the autocorrelation time is too large, and longer simulations are required to ensure reliable estimates.

In our simulations, we have used the aforementioned tests to ensure that the autocorrelation time is significantly shorter than $N_{updates}$ in each case.
Tables~\ref{tab:params_2S+1=3} and~\ref{tab:params_2S+1=4} also show technical details pertaining to the computational effort, 
where $N_{cpus}$ is the number of parallel processes and \verb~CPU time~ is
the wall clock time in seconds~\footnote{
The specific CPUs used varied between runs, as the computations were performed on the Discovery cluster 
at the USC Center for Advanced Research Computing.
Typical CPUs included Intel Xeon E5-2640 v4, AMD EPYC 7513, Intel Xeon Silver 4116, and AMD EPYC 7542.}.

In order to quantify the computational cost, we have analysed how the wall-clock time $t$ depends on the system size $N$ 
and the inverse temperature $\beta$.
For the spin-$1/2$ systems studied previously with PMR-QMC, we typically find $t \propto N^2$ and $t \propto \beta^u$ with $1.5 \lesssim u \lesssim 2.5$, and for sufficiently large systems the exponent approaches $u \simeq 2$; this behaviour is visible in Fig.~3 of Ref.~\cite{PMRQMC2024} and in Fig.~6 of Ref.~\cite{ezzell2025universalblackboxquantummonte}.

For the high-spin Heisenberg models, using the data in Tables~\ref{tab:params_2S+1=3} and~\ref{tab:params_2S+1=4},
we observe $t \propto \beta^u$ with $1 \lesssim u \lesssim 3.5$, where both $u$ and the prefactor depend sensitively on
system size and on the tuning parameter $\alpha$ that controls the typical length of the worm updates (cf.~Sec.~\ref{subsec:QMCupdates}).
Because different values of $\alpha$ were used for different system sizes, our present data permit only a crude estimate 
of the scaling with $N$ in the high-spin case, but are consistent with an effective behaviour $t \propto N^v$ with 
$2 \lesssim v \lesssim 3.5$. 
These empirical scalings are consistent with the internal cost structure of the algorithm discussed in Sec.~\ref{subsec:QMCupdates}.

It should be noted that our proposed QMC algorithm can also be readily applied to other high-spin Heisenberg models---whether 
on the square lattice or on any other graph.

\subsection{Simulating random Hamiltonians}\label{sec:Results2}

To illustrate the versatility of our algorithm, we now present simulation results for randomly generated Hamiltonians. 
Specifically, we construct random $n$-spin, $m$-term, $k$-local Hamiltonians by summing $m$ randomly generated spin operator strings.
To generate a $k$-local spin operator string, we first sample $k$ distinct spin indices $i_1, \ldots, i_k$ 
from the set $\{1, \ldots, n\}$. For each selected index, we randomly choose a spin operator from the set $\{X, Y, Z\}$, 
thereby forming a product of $k$ single-spin operators acting on distinct spins.
The resulting Hamiltonian takes the form $\sum_i c^{(i)} S^{(i)}$, 
where each operator string $S^{(i)}$ is multiplied by a real-valued coefficient $c^{(i)}$ 
drawn uniformly at random from the interval $[-1, 1]$.

\begin{figure}[t]
\includegraphics[width=\columnwidth]{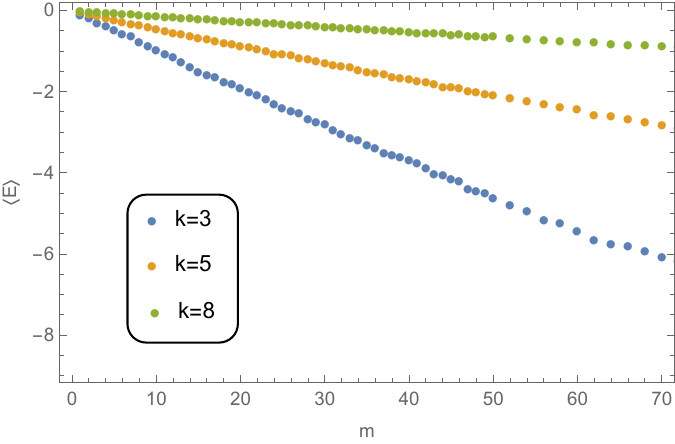}
\includegraphics[width=\columnwidth]{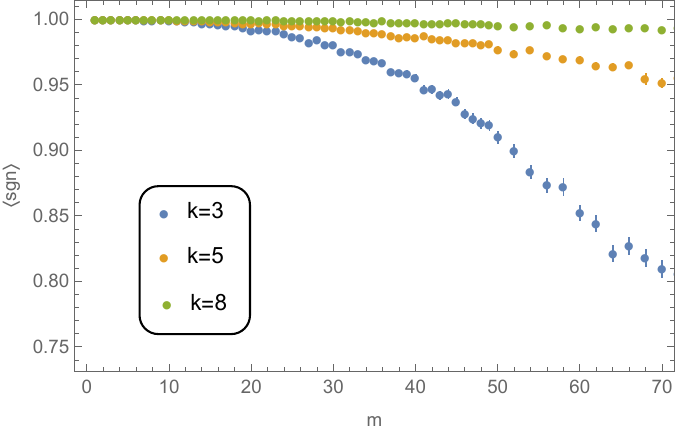}
\caption{Top: Average energy $\langle E\rangle$ over $200$ randomly generated spin-$1$ Hamiltonian instances as a function of $m$ for random $k$-local $40$-spin  Hamiltonians for $k=3$, $k=5$, and $k=8$ at $\beta=1$. Bottom: A similar plot for $\langle\sgn\rangle$, averaged over the $200$ Hamiltonian instances.}
\label{fig:randomH_2S+1=3_beta=1}
\end{figure}

\begin{figure}[t]
\includegraphics[width=\columnwidth]{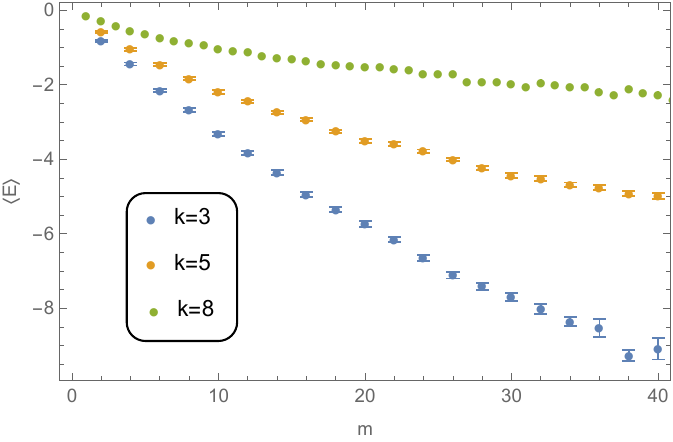}
\includegraphics[width=\columnwidth]{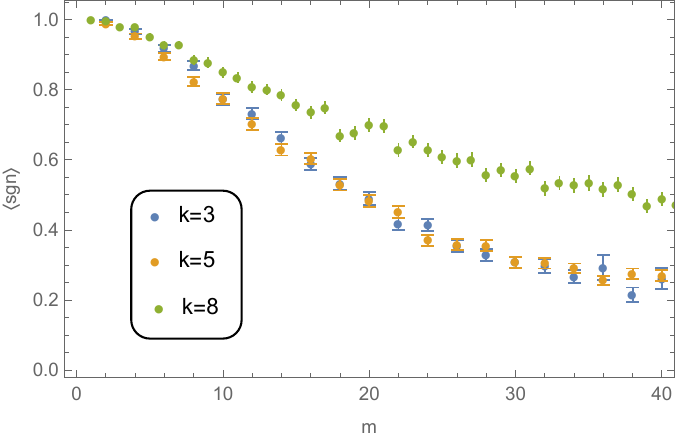}
\caption{Top: Average energy $\langle E\rangle$ over $200$ randomly generated spin-$1$ Hamiltonian instances as a function of $m$ for random $k$-local $40$-spin  Hamiltonians for $k=3$, $k=5$, and $k=8$ at $\beta=5$. Bottom: A similar plot for $\langle\sgn\rangle$, averaged over the $200$ Hamiltonian instances.}
\label{fig:randomH_2S+1=3_beta=5}
\end{figure}

\begin{figure}[t]
\includegraphics[width=\columnwidth]{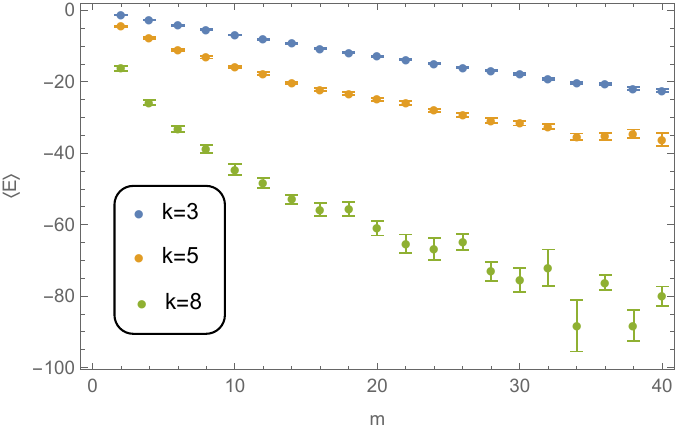}
\includegraphics[width=\columnwidth]{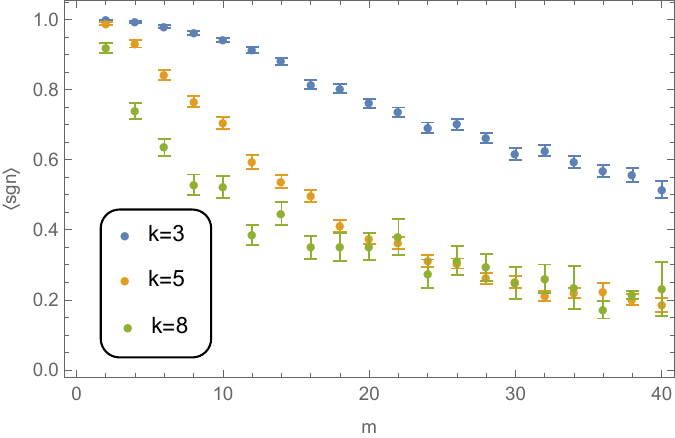}
\caption{Top: Average energy $\langle E\rangle$ over $200$ randomly generated spin-$3/2$ Hamiltonian instances as a function of $m$ for random $k$-local $40$-spin  Hamiltonians for $k=3$, $k=5$, and $k=8$ at $\beta=1$. Bottom: A similar plot for $\langle\sgn\rangle$, averaged over the $200$ Hamiltonian instances.}
\label{fig:randomH_2S+1=4_beta=1}
\end{figure}

\begin{figure}[t]
\includegraphics[width=\columnwidth]{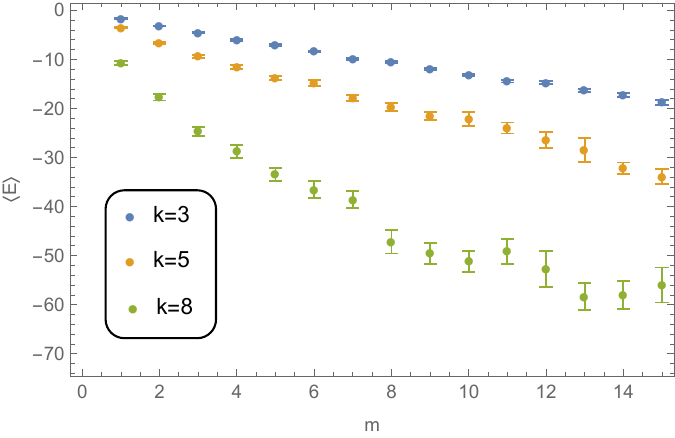}
\includegraphics[width=\columnwidth]{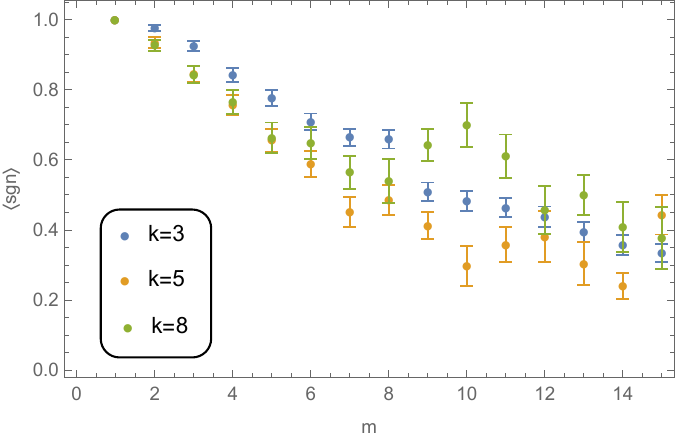}
\caption{Top: Average energy $\langle E\rangle$ over $200$ randomly generated spin-$3/2$ Hamiltonian instances as a function of $m$ for random $k$-local $40$-spin  Hamiltonians for $k=3$, $k=5$, and $k=8$ at $\beta=5$. Bottom: A similar plot for $\langle\sgn\rangle$, averaged over the $200$ Hamiltonian instances. }
\label{fig:randomH_2S+1=4_beta=5}
\end{figure}

To illustrate the ease with which our approach enables the simulation of such systems, 
we generated random $m$-term Hamiltonians for $40$ spins, sampling $200$ random instances 
for each value of $m$ and for three choices of locality: $k=3$, $k=5$, and $k=8$.
The top panels of Figs.~\ref{fig:randomH_2S+1=3_beta=1}, \ref{fig:randomH_2S+1=3_beta=5}, \ref{fig:randomH_2S+1=4_beta=1}, and \ref{fig:randomH_2S+1=4_beta=5}
display the average energy $\langle E \rangle$ over the $200$ instances as a function of $m$, 
for the cases $(s=1,\beta=1)$, $(s=1,\beta=5)$, $(s=3/2,\beta=1)$, and $(s=3/2,\beta=5)$, respectively. 
The error bars represent the magnitude of fluctuations in $\langle E \rangle$.
The bottom panels show the average sign $\langle \mathrm{sgn} \rangle$, computed over the $200$ instances 
for each combination of $m$ and $k$.

\section{Extending the framework to arbitrary mixed Hamiltonians}\label{sec:qmcext}

The method presented above can be extended to 'mixed-spin' Hamiltonians, where particles of different species 
exist and interact. To generalize the method for this case,
each particle is assigned a spin value $s_i$, rather than a single global spin value $s$ 
shared by all particles as in the previous derivations.

Moreover, incorporating fermionic and bosonic degrees of freedom can be readily accomplished, as we outline below.

\subsection{Mixed-spin Hamiltonians}

For a mixed-spin Hamiltonian, the permutation operators can be written as
\beq
P_i = \bigotimes_{j=1}^n P^{(i,j)} = \bigotimes_{j=1}^n P(2s_j+1)^{n_{i,j}},
\eeq
where $P(2s_j+1)$ is the $(2s_j+1)\times(2s_j+1)$ permutation matrix given by Eq.~(\ref{eq:Pmatrix}).
As a consequence, each permutation operator $P_i$ 
can be represented as an integer-string $p_i = [n_{i,1} n_{i,2} \cdots n_{i,n}]$,
where $n_{i,j} \in \{0, 1, \ldots, 2s_j\}$ (cf.~Sec.~\ref{sec:fundamentalcycles}).

Similar to the single species case, a multiset of operators $\{P_{i_1},\dots,P_{i_q}\}$ is represented by an integer string $[a_1 \, a_2 \, \dots \, a_M]$,
where $a_k$ denotes the number of occurrences of the operator $P_k$ among $P_{i_1}, \dots, P_{i_q}$.
In order to find fundamental cycles for the to-be-simulated Hamiltonian,
one needs to find integer strings $[a_1 \, a_2 \, \dots \, a_M]$ such that
\beq
\sum_{i=1}^M a_i \cdot n_{i,j} \equiv 0 \, \pmod{2s_j+1},\, j = 1,2,\dots,n \,. 
\label{eq:modmixed}
\eeq
For $s_1 = s_2 = \ldots = s_n = s$, the above problem reduces to Eq.~(\ref{eq:mod2s1}).
The more general case may also be solved efficiently, recalling that the system can be rewritten as Eq.~(\ref{app:dioph}), which is
a system of Diophantine equations (cf. Sec.~\ref{sec:fundamentalcycles}).

\subsection{Incorporation of fermions}

The addition of fermions to any mixed-spin model may be accomplished by converting the fermionic degrees of freedom to a  spin-$1/2$ particle representation. This is carried out via the application of a Jordan-Wigner transformation (JWT)~\cite{jwt} which maps the second-quantized annihilation operator  $c_{j }$ to an operator on $j$ spins according to
\beq
c_{j } \to\left( \prod_{k=1}^{j-1} Z_{k } \right)\frac{X_{j }-i Y_{j }}{2}
\eeq
so that \hbox{$c^{\dagger}_{j } c_{j } = (\ident+Z_{j })/2$}.
To write the Fermi-Hubbard Hamiltonian in PMR form, we rewrite the JWT as products of a diagonal operator (a function of Pauli-$Z$ strings) and a permutation operator (a Pauli-$X$): 
\begin{align}
&c_{j } \to\left[\left( \prod_{k=1}^{j-1} Z_{k } \right)\frac{\ident-Z_{j }}{2}\right]X_{j }\,,
\\
&c_{j }^\dagger \to\left[\left( \prod_{k=1}^{j-1} Z_{k } \right)\frac{\ident+Z_{j }}{2}\right] X_{j }\,.
\end{align}
Once the fermionic sites are labeled, the fermionic degrees of freedom are mapped to spin-half operators, in which case the Hamiltonian reverts to being a mixed-spin model again.

\subsection{Incorporation of bosons}

As for the inclusion of bosonic degrees of freedom, we will see in this section that the creation and annihilation operators
in the second-quantized basis correspond to permutation operators of infinite order (see also Ref.~\cite{PhysRevB.109.134519}).

As the computational basis for the PMR expansion, we use the second quantized occupation number basis for bosons, where a basis state is given as $|{\bf n}\rangle = \vert n_1,n_2,\ldots,n_L \rangle$ with $L$ being the number of sites and $n_1, \ldots, n_L$ are nonnegative integers representing the number of bosons in each site. We denote the total number of bosons, $\sum_{i=1}^L n_i$, by $n$. 
The operators $\hat{b}_{i}^{\dag}, \hat{b}_{i}$ are creation and annihilation operators, respectively, obeying
\begin{equation}
\hat{b}_{i}^{\dag} |n_1,\ldots,n_i,\ldots,n_L\rangle  = \sqrt{(n_i+1)}  |n_1,\ldots n_{i+1},\ldots,n_L\rangle \,,
\end{equation}
where $|{\bf n}^{(i,j)}\rangle$ stands for the state $|{\bf n}\rangle$ with one additional boson at site $i$ and one fewer at site $j$. The operator $\hat{n}_i=\hat{b}_{i}^{\dag} \hat{b}_{i}$ is the number operator. Define the following operator on the state in Fock space $\ket{{\bf{n}}} = \ket{n_1 , \ldots , n_i , \ldots, n_j , \ldots, n_L}$
\begin{align}
    P_m \ket{{\bf{n}}}
    &= \ket{{\bf{n}}^{(i_m , j_m)}} \notag \\
    &\equiv \ket{n_1 , \ldots , n_{i_m}+1, \ldots, n_{j_m}-1 , \ldots, n_L} \, .
\end{align}
$P_m$ can be thought of as a permutation operator, permuting the states with different number of bosons on $i_m$ and $j_m$ site. It is important to note that application of $b^{\dagger}_{i_m} b_{j_m}$ would annihilate a state with $n_{j_m} = 0$. However, one can allow $P_m$ to simply map states with $n_{j_m} = 0$ to states with negative $n_{j_m}$. The annihilation will be imposed by the matrices $D_m$.

Given this construction, for every $P_m$ one can write a corresponding inverse permutation $(P_m)^{-1}$ that reverses the mapping. Moreover, $P_m$ loses its finite periodicity, as there is no longer any finite $s$ such that $P_m^{2s+1} = \ident$. Thus, the permutation operators $P_m$ have infinite order. In this case, finding identity equivalent string of permutations will be tantamount to finding the mod-$\infty$ nullspace of vector of integer. This is nothing but solving a system of linear equations with integer solutions, i.e. a system of Diophantine equations.

Given a set of bosonic permutations $\{P_i\}$, we can equivalently use integer strings $\{\boldsymbol{p}_i\}$ to denote the action of a particular $P_i$ on the a Fock space basis state via an integer string $\boldsymbol{p}_i$. In this way, in order to find all identity equivalent string $\Pi_i P_i = \ident$, we will need to find the set of solutions to the following equation
\begin{align}
    \sum_i a_i \boldsymbol{p}_i = \boldsymbol{0} \, ,
\label{eq:bosons_dioph}
\end{align}
where $a_i$ are unknown integers. 
Equation~(\ref{eq:bosons_dioph}) can be addressed using standard methods for Diophantine equations.

\section{Summary and discussion}\label{sec:conclusions}
We presented a universal Trotter-error-free quantum Monte Carlo scheme capable of simulating, for the first time, arbitrary high-spin Hamiltonians. We have demonstrated that the permutation matrix representation of Hamiltonians allows one to automatically produce QMC updates that are provably ergodic and satisfy detailed balance, thereby ensuring the convergence of the Markov chain to the equilibrium. 

Our algorithm therefore enables one to study the equilibrium properties of essentially any conceivable high-spin system using a single piece of code. While our approach guarantees a correct equilibrium distribution of the Markov chain, the algorithm does not guarantee a universal rapid mixing of the Markov chain, nor does it resolve or aim to resolve the sign problem. 

We illustrated this ability by producing results for the quantum Heisenberg model for two types of spin particles, namely, spin-$1$ and spin-$3/2$. To demonstrate the versatility of our approach, we have studied in addition the equilibrium properties of randomly generated Hamiltonians, which existing QMC techniques cannot simulate. 

We have further shown that the methodologies devised here can be extended to other particle species and mixtures thereof.
We hope that the feasibility of such calculations and applicability of our code to a wide range of high-spin systems
will contribute to the broader understanding of quantum magnetism and phase transitions in low-dimensional quantum systems.

We believe that the generality and versatility of the method developed here will make our proposed technique a very useful tool for condensed matter physicists studying spin systems, allowing the community to explore with ease physical models that have so far been inaccessible, cumbersome to code, or too large to implement with existing techniques. To that aim, we have made our program code freely accessible on GitHub~\cite{github_PMRQMC_high_spin}.


\vspace{0.5cm}
\begin{acknowledgments}
This project was supported in part by NSF award \#2210374.
The authors acknowledge the Center for Advanced Research Computing (CARC) at the University of Southern California for providing the computing resources used in this work.
In addition, this material is based upon work supported by the Defense Advanced Research Projects Agency (DARPA) under Contract No. HR001122C0063. All material, except scientific articles or papers published in scientific journals, must, in addition to any notices or disclaimers by the Contractor, also contain the following disclaimer: Any opinions, findings and conclusions or recommendations expressed in this material are those of the author(s) and do not necessarily reflect the views of the Defense Advanced Research Projects Agency (DARPA). 

\end{acknowledgments}
\bibliography{refs}

\appendix

\section{Cycle completion with gaps}
\label{appendix:cycle_completion}

As was mentioned in the main text, a simple cycle completion protocol, in which consecutive elements of $S_{{\bf{i}}_q}$
form a subsequence $S$, may lead to a violation of the ergodicity condition and therefore also an incorrect calculation.
The reason for this is the possibility of the resultant configuration having zero weight [as per Eq.~(\ref{eq:gbw})]. As a consequence, the fundamental cycle completion routine may never be applied for some of the fundamental cycles
during the Markov process, which can in turn lead to the ergodicity violation.

To resolve this issue, we have developed a protocol that we call `cycle completion with gaps'. This protocol does not require the elements of the sequence $S$ to form a consecutive unit within $S_{{\bf{i}}_q}$.
A detailed description of this subroutine follows.

The parameters are: $r_{\min}$, $r_{\max}$, $l_{\min}(r)$, $l_{\max}(r)$.
One can choose: $r_{\min} = (f_{\min} - 1)/2$, $r_{\max} = (f_{\max} + 1)/2$, 
$l_{\min}(r)=2r-1$, $l_{\max}(r) = 2r+1$,
where $f_{\min}$ and $f_{\max}$ are minimal and maximal fundamental cycles lengths, respectively.
The other reasonable option (`exhaustive search') is to choose $r_{\min} = 0$, $r_{\max} = f_{\max}$,
$l_{\min}(r) = r$, $l_{\max}(r) = f_{\max}$.

The sequence of operations is as follows.
\begin{enumerate}
\item Pick a random integer $u$ according to a geometric distribution $p_u$. As we'll see, $u$ is the total number of operators in the `gaps'.
\item If $q < u + r_{\min}$, then the update is rejected.
\item Pick a random integer $r$ such that $r_{\min} \leq r \leq \min(r_{\max},q-u)$.
      We note that the probability $p_r(q) = (\min(r_{\max},q-u)-r_{\min}+1)^{-1}$ depends on $q$.
\item Randomly pick a sub-sequence $\widetilde{S}$ of length $r+u$ containing consecutive operators from the sequence $S_{{\bf i}_q}$.
\item Randomly choose a subsequence $S$ of length $r$ from $\widetilde{S}$. The remaining $u$ operators in $\widetilde{S}$ we will call `gaps'.
\item With probability $1/2$ we set the {\it inv} flag to {\it true}, otherwise to {\it false}.
\item Find all fundamental cycles of lengths $l$ such that $l_{\min}(r) \leq l \leq l_{\max}(r)$, each containing all operators 
      of $S$ if ${\text\it inv} = \text{\it false}$, or all operators of $S^{-1}$ if ${\text\it inv} = \text{\it true}$.
      Here, $S^{-1}$ denotes the sequence in which each operator from $S$ is replaced by its inverse operator.
      Denote by $n_c$ the number of found cycles.
\item If $n_c = 0$, the update is rejected. Otherwise, we randomly choose one of the found fundamental cycles.
      Let us denote by $S'$ the sequence consisting of all the remaining $r'$ operators from the selected cycle.
\item Attempt to replace the sub-sequence $\widetilde{S}$ of length $r+u$ by the sequence $\widetilde{S'}$ of length $r'+u$.
      Here, $\widetilde{S}$ contains all operators of $S'$, as well as all the `gaps' if ${\text\it inv} = \text{\it true}$,
      and otherwise it contains all operators of $S'^{-1}$, as well as all the `gaps'.
      We shuffle the sequence $\widetilde{S'}$ so that its operators are contained in random order.
      We accept the update with the probability $P_{accept}$, which is considered below.
\end{enumerate}

Compared to the spin-1/2 version of this algorithm outlined in Ref.~\cite{PMRQMC2024}, 
here the sequence $S$ may contain repetitions.
Let the sequence $S$ contain $s_i$ operators $P_i$ such that $\sum_i s_i = r$,
and the sequence $S'$ contain $s'_i$ operators $P_i$ such that $\sum_i s'_i = r'$.
Let us find the acceptance probability $P_{accept}$ such that the detailed balance holds for the above protocol.
Suppose that the $u$ gaps contain $u_i$ of operators $P_i$, where $i = 1,2,\dots, M$, so that $\sum_i u_i = u$.
Let us denote the old and new configurations as $A$ and $B$, probability to select $B$ from $A$ as $P_{select}(A\to B)$,
probability to select $A$ from $B$ as $P_{select}(B\to A)$. Then, we have
\begin{multline}
P_{select}(A\to B) = p_u \cdot p_r(q) \cdot (q-(r+u)+1)^{-1} \frac{1}{n_c} \times \\ \times \binom{r+u}{u}^{-1} \binom{r'+u}{u}^{-1} \cdot \frac{s'_1!\dots s'_M!}{r'!} \cdot \frac{u_1! \dots u_M!}{u!},
\end{multline}
\begin{multline}
P_{select}(B\to A) = p_u \cdot p_r(q') \cdot (q'-(r'+u)+1)^{-1} \frac{1}{n'_c} \times \\ \times \binom{r'+u}{u}^{-1} \binom{r+u}{u}^{-1} \cdot \frac{s_1!\dots s_M!}{r!} \cdot \frac{u_1! \dots u_M!}{u!}.
\end{multline}
Here, $n'_c$ is the number of fundamental cycles of lengths $l$ such that $l_{\min}(r') \leq l \leq l_{\max}(r')$, each containing 
all $r'$ operators of the sub-sequence $S'$.
Since $q'=q+r'-r$, we have $q-(r+u)+1 = q'-(r'+u)+1$. Therefore,
\begin{align}
P_{accept}(A\to B) = \min\left(1,\frac{W_B}{W_A}\cdot\frac{P_{select}(B\to A)}{P_{select}(A\to B)}\right) \notag \\
=\min\left(1,\frac{W_B}{W_A}\cdot\frac{p_{r}(q')}{p_r(q)}\cdot\frac{n_c}{n'_c}\cdot
\frac{r'!}{s'_1\dots s'_M}\cdot\frac{s_1!\dots s_M!}{r!}\right).
\label{eq:Paccept}
\end{align}
Here, $W_A$ and $W_B$ are the weights of the old and the new operator sequences.
Because $P(A\to B) = P_{select}(A\to B) P_{accept}(A\to B)$
and $P(B\to A) = P_{select}(B\to A) P_{accept}(B\to A)$, Eq.~(\ref{eq:Paccept}) satisfies the detailed balance condition.
\vspace{0.8cm}

\end{document}